\documentclass[12pt]{article}
\usepackage{epsfig}
\usepackage{color}
\usepackage{amssymb,amsmath}
\usepackage{graphicx}
\usepackage{here}
\usepackage{epsfig}

\newcommand{\ba}{\begin{align}}
\newcommand{\ea}{\end{align}}

\def\tr{\mathrm{tr}}
 \makeatletter
\def\alt{\mathrel{\mathpalette\gl@align<}}
\def\agt{\mathrel{\mathpalette\gl@align>}}
\def\gl@align#1#2{\lower.6ex\vbox{\baselineskip\z@skip\lineskip\z@
\ialign{$\m@th#1\hfil##\hfil$\crcr#2\crcr\sim\crcr}}} \makeatother

\begin{document}
\begin{flushright}
\end{flushright}
\vspace*{1.0cm}

\begin{center}
\baselineskip 20pt 
{\Large\bf 
Proton Decay Prediction in 5D Gauge-Higgs Unification
}
\vspace{1cm}

{\large 
Naoyuki Haba$^a$, \ Nobuchika Okada$^b$ \ and \ Toshifumi Yamada$^a$
} \vspace{.5cm}

{\baselineskip 20pt \it
$^a$ Graduate School of Science and Engineering, Shimane University, Matsue 690-8504, Japan
$^b$ Department of Physics and Astronomy, University of Alabama, \\ Tuscaloosa, Alabama 35487, USA
}

\vspace{.5cm}

\vspace{1.5cm} {\bf Abstract} \end{center}

The Higgs boson mass and top quark mass imply that the Higgs quartic coupling vanishes around the scale of $10^9 - 10^{13}$~GeV, depending on the precise value of the top quark mass. 
The vanishing quartic coupling can be naturally addressed if the Higgs field originates from a 5-dimensional gauge field
 and the 5th dimension is compactified at the scale of the vanishing Higgs quartic coupling, which is a scenario based on gauge-Higgs unification.
We present a general prediction of the scenario on the proton decay process $p \to \pi^0 e^+$.
In many gauge-Higgs unification models, the 1st generation fermions are localized towards an orbifold fixed point in order to realize the realistic Yukawa couplings.
Hence, four-fermion operators responsible for the proton decay can appear with a suppression of the 5-dimensional Planck scale (not the 4-dimensional Planck scale). 
Since the 5-dimensional Planck scale is connected to the compactification scale,
 we have a correlation between the proton partial decay width and the top quark mass. 
We show that the future Hyper-Kamiokande experiment may discover the proton decay if the top quark pole mass is larger than about 172.5~GeV.

\thispagestyle{empty}

%\bigskip
\newpage

%\addtocounter{page}{-1}
\setcounter{footnote}{0}
%%%%%%%%%%%%%%%%%%%%%%%%%%
%\baselineskip 36pt
% Main body
%%%%%%%%%%%%%%%%%%%%%%%%%%
\baselineskip 18pt
%%%%%%%%%%%%%%%%%%%%%%%%%%
%

The determination of the Higgs boson mass at $m_h = 125.09\pm0.24$~GeV~\cite{higgs}, together with the top quark mass measurment~\cite{topatlas,topcms},
 has introduced a new energy scale to the Standard Model (SM): 
 the scale at which the Higgs field quartic coupling vanishes through its renormalization group (RG) running, 
 hereafter denoted by $\Lambda_{cr}$,
 which is located about $10^9 - 10^{13}$~GeV depending sensitively on the top quark mass.
The SM can remain viable above the scale $\Lambda_{cr}$, since the Universe is sufficiently long-lived even if the Higgs quartic coupling turns negative
 above $\Lambda_{cr}$~\cite{lifetime}.
However, the scale $\Lambda_{cr}$ may indicate some new physics beyond the SM, in which
 the Higgs quartic coupling vanishes above $\Lambda_{cr}$, and below $\Lambda_{cr}$, the theory is effectively described by the SM 
 where the RG running induces a non-zero Higgs quartic coupling.

Gauge-Higgs unification~\cite{ghu} in a 5-dimensional (5D) Minkowski spacetime generally predicts the vanishing of the Higgs quartic coupling above the compactification scale of the 5th dimension, namely, the Kaluza-Klein (KK) scale.
This is because the Higgs field is embedded in the 5th dimensional component of a gauge field
 and the gauge symmetry forbids a tree-level potential for the Higgs field.
The gauge symmetry is explicitly broken in an orbifold compactification of the 5th dimension
 and the resultant KK modes of gauge fields and bulk fermions induce the Higgs potential radiatively.
By matching the effective potential generated by the tower of KK modes with that generated by the zero mode,
 Ref.~\cite{eft} has proved the so-called "gauge-Higgs condition", which states that
 the Higgs quartic coupling vanishes at the KK scale in general gauge-Higgs unification models.
Hence $\Lambda_{cr}$ of the SM may suggest the KK scale of a gauge-Higgs unification model.

As a common prediction of 5D gauge-Higgs unification models where $\Lambda_{cr}$ of the SM corresponds to the KK scale,
 we focus on the proton decay process $p \to \pi^0 e^+$ induced by Planck-suppressed operators.
At the orbifold fixed points, quantum gravity can induce four-fermion operators suppressed by the Planck scale of the 5D spacetime, $M_5$.
Since fermions in the 5D spacetime couple with the Higgs field with the strength of the weak gauge coupling,
 the SM 1st generation quarks and leptons are necessarily localized towards an orbifold fixed point to avoid too large Yukawa couplings.
Hence four-fermion operators involving the 1st generation fermions, which are responsible for the $p \to \pi^0 e^+$ process,
 naturally arise with a factor of $1/M_5^2$.
This is in contrast with fermions that reside totally in the bulk, for which, after integrating over the 5th dimension,
 four-fermion operators arise with a factor of $1/M_4^2$ in the 4-dimensional (4D) effective theory,
 where $M_4 \simeq 2.44\times 10^{18}$~GeV is the reduced Planck mass of the 4D spacetime.
$M_5$ is tied with the compactification scale $L$ by $M_5^3L = M_4^2$ and hence with the KK scale $\sim 1/L$.
We thus find a correlation between $\Lambda_{cr}$ of the SM and the partial decay width for the $p \to \pi^0 e^+$ process.
Furthermore, since $\Lambda_{cr}$ is sensitive to the top quark mass, the above correlation is translated into that between the top quark mass and the proton decay rate,
 which we will present in this letter.

The above correlation holds in general models of gauge-Higgs unification provided the 1st generation fermions are localized towards an orbifold fixed point.
In this letter, however, we first present a concrete model of gauge-Higgs unification where the 1st generation matter is localized,
 to prove that such models exist, and then work in this particular model to illustrate how the correlation is derived.
For this purpose, we consider the minimal setup for gauge-Higgs unification, which is similar to models in Refs.~\cite{scrucca,csaki}.
The model is based on a 5D flat spacetime compactified on $S^1/Z_2$ and contains $SU(3)_w \times U(1)_v$ gauge group that is explicitly broken into $SU(2)_L \times U(1)_Y$ at the orbifold fixed points.
The massless component of the 5th dimensional $SU(3)_w$ gauge field is identified with the SM Higgs field.
In the setup, the simplest mechanism is adopted to derive the SM Yukawa couplings.
We introduce 4D Weyl fermions localized at an orbifold fixed point, and bulk Dirac fermions in the 5D spacetime,
 whose left or right-handed components satisfy Neumann condition at the orbifold fixed point and mix with the localized fermions through 4D Dirac mass terms.
The SM fermions are given as mixtures of the 4D and 5D fermions, and their couplings with the Higgs field are controlled by the 4D Dirac mass.
We further introduce 4D localized operators involving four 4D fermions at the orbifold fixed point suppressed by the 5D Planck scale $M_5$,
 which are responsible for the proton decay.

This letter is organized as follows:
We first describe the minimal setup for gauge-Higgs unification with emphasis on the fermion sector.
Next we review the effective theory approach to gauge-Higgs unification studied in Ref.~\cite{eft},
 and derive the relation between $\Lambda_{cr}$ and the KK scale.
We then introduce 5D Planck-suppressed operators that induce the proton decay.
Eventually, we derive a correlation between $\Lambda_{cr}$ and the partial width of the $p \to e^+ \pi^0$ process,
 and present a plot of the top quark pole mass versus the proton partial decay width.

We present the minimal setup for gauge-Higgs unification.
However, the following argument can be extended to general models of gauge-Higgs unification.
Note that since the KK scale is as high as $10^9 - 10^{13}$~GeV, no experimental constraints other than the proton decay rate apply to the setup.
We consider a 5D flat spacetime whose 5th dimension is compactified on the orbifold $S^1/Z_2$.
The 5th dimension is parametrized by $y$ in the range $\pi R \geq y \geq -\pi R$ with the points of $y=\pi R$ and $y=-\pi R$ identified. 
The orbifolding identifies $y$ with $-y$, which gives the orbifold fixed points at $y=0,\pi R$.
In the bulk, we introduce $SU(3)_C\times SU(3)_w\times U(1)_v$ gauge group, where $SU(3)_C$ is the color in the SM.

We demand that the 4D and 5D components of the $SU(3)_w$ gauge field $(w_{\mu},w_5)$ and those of $U(1)_v$ gauge field $(v_{\mu},v_5)$ transform under the orbifolding as
\begin{align}
w_{\mu}(y) &= P^{\dagger}w_{\mu}(-y)P, \ \ \ w_5(y) = -P^{\dagger}w_5(-y)P, 
\ \ \ v_{\mu}(y) = v_{\mu}(-y), \ \ \ v_5(y) = -v_5(-y) \nonumber \\
{\rm with} \ P &= \left(
\begin{array}{ccc}
-1 & 0& 0 \\
0 & -1& 0 \\
0 &  0& 1
\end{array}
\right),
\label{orb}
\end{align}
 where $P$ acts in the $SU(3)_w$ gauge space.
It follows that the boundary conditions at $y=0,\pi R$ explicitly break $SU(3)_w$ into $SU(2)_L \times U(1)_X$ and 
 accordingly the gauge boson is decomposed as $\mathbf{8} \to \mathbf{3}_0 + \mathbf{2}_{\sqrt{3}/2} + \mathbf{2}_{-\sqrt{3}/2} + \mathbf{1}_0$, where the subscripts denote $U(1)_X$ charge.
$\mathbf{3}_0+\mathbf{1}_0$ of $w_{\mu}$, $\mathbf{2}_{\sqrt{3}/2}+\mathbf{2}_{-\sqrt{3}/2}$ of $w_5$ and $v_{\mu}$ satisfy Neumann conditions at $y=0,\pi R$ and thus have the zero mode in the KK expansion,
 while the rest of the gauge fields satisfy Dirichlet conditions and have no zero mode.
We identify $SU(2)_L$ with the SM weak gauge group and $\mathbf{2}_{\sqrt{3}/2}+\mathbf{2}_{-\sqrt{3}/2}$ of $w_5$ with the SM Higgs field,
 and further assume that $U(1)_X \times U(1)_v$ breaks into the SM hypercharge $U(1)_Y$ leading to the correct Weinberg angle.

In the bulk, we introduce three copies of 5D Dirac fermions $\Psi$'s in $(\mathbf{3}, \mathbf{3})$, $(\mathbf{3}, \mathbf{\bar{6}})$ and $(\mathbf{3}, \mathbf{10})$ representations of $SU(3)_C\times SU(3)_w$ with no $U(1)_v$ charge
 (they are in the fundamental, symmetric and rank-three symmetric representations of the $SU(3)_w$),
 and thier partners $\tilde{\Psi}$'s with the same gauge charge.
We will see that the only role of $\tilde{\Psi}$ is to allow $Z_2$ invariant Dirac mass term between $\Psi$ and $\tilde{\Psi}$ which turns the KK zero modes of $\Psi$ and $\tilde{\Psi}$ to be massive 
 and makes the model phenomenologically viable.
The bulk fermions always transform under the orbifolding as $\bar{\Psi}\Psi(y)=-\bar{\Psi}\Psi(-y)$, $\bar{\tilde{\Psi}}\tilde{\Psi}(y)=-\bar{\tilde{\Psi}}\tilde{\Psi}(-y)$.
We impose the following boundary conditions:
\begin{align}
\Psi(y=0) &= -\gamma_5 R(P) \Psi(y=0), \ \ \ \Psi(y=\pi R) = -\gamma_5 R(P) \Psi(y=\pi R), \nonumber \\
\tilde{\Psi}(y=0) &= \gamma_5 R(P) \tilde{\Psi}(y=0), \ \ \ \tilde{\Psi}(y=\pi R) = \gamma_5 R(P) \tilde{\Psi}(y=\pi R),
\end{align}
 where $R(P)$ denotes $P$ in the representation of $SU(3)_w$ to which $\Psi$ and $\tilde{\Psi}$ belong.
Along the breaking of $SU(3)_w \to SU(2)_L \times U(1)_X$ at $y=0,\pi R$, each representation of $SU(3)_w$ is decomposed as
 $\mathbf{3} \to \mathbf{2}_{1/2\sqrt{3}} + \mathbf{1}_{-1/\sqrt{3}}$, 
 $\mathbf{\bar{6}} \to \mathbf{3}_{-1/\sqrt{3}} + \mathbf{2}_{1/2\sqrt{3}} + \mathbf{1}_{2/\sqrt{3}}$
 and $\mathbf{10} \to \mathbf{4}_{\sqrt{3}/2} + \mathbf{3}_0 + \mathbf{2}_{-\sqrt{3}/2} + \mathbf{1}_{-\sqrt{3}}$.
Among the components of $\Psi$, the right-handed components of the two $(\mathbf{3}, \mathbf{2})_{1/2\sqrt{3}}$'s, $(\mathbf{1}, \mathbf{4})_{\sqrt{3}/2}$ and $(\mathbf{1}, \mathbf{2})_{-\sqrt{3}/2}$ 
 and the left-handed components of $(\mathbf{3}, \mathbf{1})_{-1/\sqrt{3}}$, $(\mathbf{3}, \mathbf{3})_{-1/\sqrt{3}}$, $(\mathbf{3}, \mathbf{1})_{2/\sqrt{3}}$, $(\mathbf{1}, \mathbf{3})_0$ and $(\mathbf{1}, \mathbf{1})_{-\sqrt{3}}$
 (each bracket denotes the $SU(3)_C \times SU(2)_L$ charge and each subscript the $U(1)_X$ charge)
 satisfy Neumann condition at the boundaries.
As to $\tilde{\Psi}$, the same gauge components with the opposite chirality satisfy Neumann condition.

At the orbifold fixed points, the gauge symmetry is $SU(3)_C \times SU(2)_L \times U(1)_X \times U(1)_v$.
At $y=0$, we introduce three copies of 4D localized left-handed Weyl fermions $\chi$'s in $(\mathbf{3}, \mathbf{2})_{1/2\sqrt{3}}$ and $(\mathbf{1}, \mathbf{2})_{-\sqrt{3}/2}$ representations
 and right-handed Weyl fermions $\tilde{\chi}$'s in $(\mathbf{3}, \mathbf{1})_{-1/\sqrt{3}}$, $(\mathbf{3}, \mathbf{1})_{2/\sqrt{3}}$ and $(\mathbf{1}, \mathbf{1})_{-\sqrt{3}}$ representations
 of the $SU(3)_C \times SU(2)_L \times U(1)_X$ gauge group, without $U(1)_v$ charge.
They exactly correspond to the SM fermions.
They have 4D Dirac mass terms with 
 the right-handed components of the two $(\mathbf{3}, \mathbf{2})_{1/2\sqrt{3}}$'s and $(\mathbf{1}, \mathbf{2})_{-\sqrt{3}/2}$
 and the left-handed components of $(\mathbf{3}, \mathbf{1})_{-1/\sqrt{3}}$, $(\mathbf{3}, \mathbf{1})_{2/\sqrt{3}}$ and $(\mathbf{1}, \mathbf{1})_{-\sqrt{3}}$ of $\Psi$'s,
 since they satisfy Neumann condition.
On the other hand, the SM fermions do not couple with any components of $\tilde{\Psi}$'s.

With the field content above, the action is schematically written as
\begin{align}
S &= \int {\rm d}^4x \int_{-\pi R}^{\pi R} {\rm d}y \, \left[ \, \frac{1}{2}M_5^3 {\cal R} -\frac{1}{2}{\tr}[w_{MN}w^{MN}] -\frac{1}{4}v_{MN}v^{MN} \right. \nonumber \\
&+ i\bar{\Psi} \gamma^M D_M \Psi + i\bar{\tilde{\Psi}} \gamma^M D_M \tilde{\Psi} - \hat{M}\bar{\Psi}\tilde{\Psi} - {\rm h.c.} \nonumber \\
&+ \left. \delta(y) \left( i\bar{\chi} \sigma^{\mu} D_{\mu} \chi + i\bar{\tilde{\chi}} \bar{\sigma}^{\mu} D_{\mu} \tilde{\chi} + m_1\bar{\Psi}_R \chi + m_2\bar{\Psi}_L \tilde{\chi} + {\rm h.c.} \right) \, \right]
\label{action}
\end{align}
 where $M,N=0,1,2,3,5$ are 5D spacetime indices,
 and $w_{MN}$ and $v_{MN}$ denote the field strength of the $SU(3)_w$ gauge field $(w_{\mu},w_5)$ and the $U(1)_v$ gauge field $(v_{\mu},v_5)$, respectively.
$\hat{M}$ denotes $Z_2$ invariant 5D Dirac mass for the bulk fermions, which gives Dirac mass to all the KK modes including the zero mode.
The second line represents the Lagrangian localized at $y=0$, in which $\Psi_R, \Psi_L$ denote the components of $\Psi$ that satisfy Neumann condition at $y=0$
 and $m_1,m_2$ denote Dirac mass terms between them and the 4D localized fermions.
We write the massless mode of $\mathbf{2}_{\sqrt{3}/2}+\mathbf{2}_{-\sqrt{3}/2}$ component of $w_5$, which we identify with the SM Higgs field, as $H$. 
Then the action contains the following term:
\begin{align}
S &\supset \int {\rm d}^4x \, 2\pi R \left[ \, ig_5 (\bar{\Psi}_L H \Psi_R - \bar{\Psi}_R H^{\dagger} \Psi_L) + m_1\bar{\Psi}_R \chi + m_2\bar{\Psi}_L \tilde{\chi} + {\rm h.c.} \, \right],
\end{align}
 from which we obtain the SM Yukawa coupling $\bar{\tilde{\chi}} H \chi + {\rm h.c.}$ after integrating out $\Psi_R, \Psi_L$.

In Eq.~(\ref{action}), ${\cal R}$ denotes the scalar curvature and $M_5$ the 5D Planck mass, which is related to the 4D reduced Planck mass $M_4\simeq2.44\times10^{18}$~GeV as
\begin{align}
2\pi R M_5^3 &= M_4^2.
\label{planck}
\end{align}
\\

The potential for the Higgs field $H$ is zero at tree level because it is a component of the gauge field $w_5$.
The potential is generated through radiative corrections from KK modes of the gauge bosons and bulk fermions.
Ref.~\cite{eft} has investigated the general model of gauge-Higgs unification and has proved that,
 if the effective potential for the Higgs field is induced by bulk fermions satisfying Neumann condition at both boundaries,
 the running Higgs quartic coupling constant $\lambda(\mu)$ should fulfill the following condition at the scale $1/(2\pi R)$:
\begin{align}
\lambda\left( \frac{1}{2\pi R} \right) &= 0,
\label{criticality}
\end{align}
 which remains true even when the bulk fermions obtain Dirac mass below the KK scale $1/R$ from a $Z_2$ invariant 5D Dirac mass term.
The above statement applies to our setup as long as we take $\hat{M}$ in Eq.~(\ref{action}) below $1/R$,
 since some components of $\Psi$, $\tilde{\Psi}$ that satisfy Neumann condition at $y=0,\pi R$ are responsible for generating the Higgs potential.
Then the scale at which the Higgs quartic coupling vanishes, $\Lambda_{cr}$, coincides with $1/2\pi$ times the KK scale $1/R$.

We introduce 5D Planck-suppressed operators localized at the orbifold fixed point $y=0$.
The 1st generation quarks and leptons are mostly composed of 4D fermions localized at $y=0$, namely, the corresponding 4D Dirac mass terms $m_1,m_2$ in Eq.~(\ref{action}) are small,
 because the 1st generation fermions have tiny couplings with the Higgs field $H$ which is the $\mathbf{2}_{\sqrt{3}/2}+\mathbf{2}_{-\sqrt{3}/2}$ component of $w_5$.
Hence, we can generally introduce four-fermion operators among them, which are naturally suppressed by the 5D Planck scale, and in particular, we have
\begin{align}
\Delta S &= \int {\rm d}^4x \, 
\left( \, \frac{h_1}{M_5^2} \epsilon_{ab}\epsilon_{cd} (q^a q^b)(q^c \ell^d) 
 + \frac{h_2}{M_5^2} \epsilon_{ab}\epsilon_{cd} (q^a q^c)(q^d \ell^b) \right. \nonumber \\
&+ \left. \frac{h_3}{M_5^2} \epsilon_{ab} (q^a q^b)(u e)
 + \frac{h_4}{M_5^2} \epsilon_{ab} (u d)(q^a \ell^b)
 + \frac{h_5}{M_5^2} (u d)(u e) \, \right)
\label{plancksuppressed}
\end{align}
 where $q,u,d,\ell,e$ are the first generation SM fermions, $h_1,h_2,h_3,h_4,h_5$ are $O(1)$ coupling constants, $a,b,c,d$ are isospin indices
 and we take a spinor product inside each parenthesis. Here the contraction of color indices is obvious.
The partial width of the $p \to \pi^0 e^+$ process is given by~\cite{nath}
\begin{align}
\Gamma(p \to \pi^0 e^+) &= \frac{(m_p^2-m_{\pi^0}^2)^2}{64\pi f_{\pi}^2 m_p^3}(1+D+F)^2
\left\{ \left\vert \beta\frac{h_1}{M_5^2} + \beta\frac{h_2}{M_5^2} + \alpha\frac{h_4}{M_5^2} \right\vert^2 + \left\vert \alpha\frac{h_3}{M_5^2} + \beta\frac{h_5}{M_5^2} \right\vert^2 \right\}
\label{protondecay}
\end{align}
 where $\alpha$ and $\beta$ parametrize the matrix elements for three-quark operators between the vacuum and the one proton state, and 
 $D$ and $F$ are parameters of the chiral Lagrangian.
\footnote{
The effects of RG running on the operators are absorbed into the definition of $h_1,h_2,h_3,h_4,h_5$, which remain to be $O(1)$.
}

From Eqs.~(\ref{planck}),~(\ref{criticality}) and (\ref{protondecay}),
 we obtain the following relation between the scale $\Lambda_{cr}$ at which the Higgs quartic coupling vanishes, and 
 the proton decay partial width $\Gamma(p \to \pi^0 e^+)$:
\begin{align}
\Gamma(p \to \pi^0 e^+) &= \frac{(m_p^2-m_{\pi^0}^2)^2}{64\pi f_{\pi}^2 m_p^3}(1+D+F)^2 (\vert \beta h_1 + \beta h_2 + \alpha h_4 \vert^2 + \vert \alpha h_3 + \beta h_5 \vert^2) \, 
\left( \frac{1}{M_4^2 \Lambda_{cr}} \right)^{4/3}.
\end{align}
On the other hand, $\Lambda_{cr}$ sensitively depends on the top quark mass,
 whose connection to $\Lambda_{cr}$ can be evaluated by solving the RG equations for the Higgs quartic coupling.
In our setup, the massive KK modes of the gauge bosons and bulk fermions have mass equal to or above $1/R$.
Additionally, we assume that the $Z_2$ invariant Dirac mass $\hat{M}$ in Eq.~(\ref{action}) pushes the mass of the KK zero mode of the bulk fermions above $1/(2\pi R)$.
Then the field content below the scale $1/(2\pi R)$ is identical with the SM one,
 and hence we may use the SM RG equations to evaluate $\Lambda_{cr}$, as it equals to $1/(2\pi R)$.

\begin{figure}[H]
  \begin{center}
    \includegraphics[width=120mm]{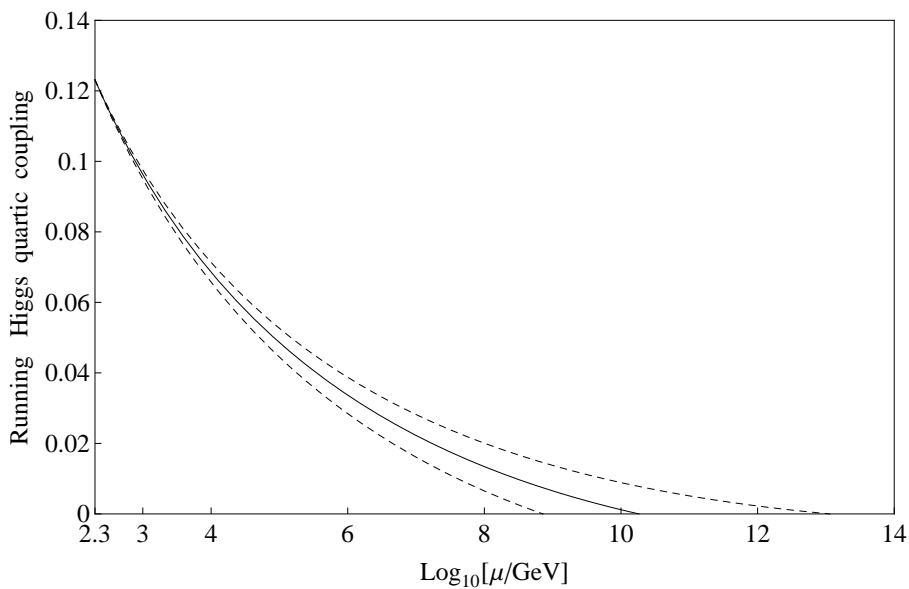}
    \caption{RG running of the Higgs quartic coupling in the SM,
     for three cases where the top quark pole mass is given by $m_t^{{\rm pole}}=171.44$~GeV (upper dashed line), $172.84$~GeV (middle solid line)
     and $174.24$~GeV (lower dashed line). 
    The parameters other than the top quark mass are fixed as $m_h=125.09$~GeV, $M_W=80.384$~GeV and $\alpha_S(M_Z)=0.1184$.
    }
    \label{rge}
  \end{center}
\end{figure}

\begin{figure}[H]
  \begin{center}
    \includegraphics[width=120mm]{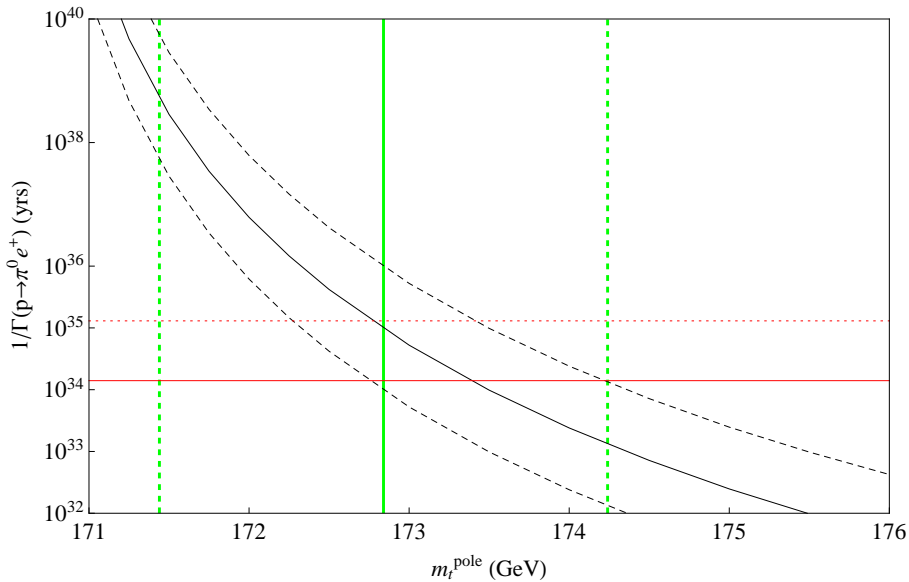}
    \caption{The correlation between the top quark pole mass $m_t^{{\rm pole}}$ and the inverse of the proton decay partial width $1/\Gamma(p \to \pi^0 e^+)$.
     The factor $(\vert \beta h_1 + \beta h_2 + \alpha h_4 \vert^2 + \vert \alpha h_3 + \beta h_5 \vert^2)$ in Eq.~\ref{protondecay} is variated from
      10$\vert\alpha\vert^2$ to 0.1$\vert\alpha\vert^2$ with $\vert\alpha\vert=0.009$~GeV$^3$, 
      and the lower dashed, solid and upper dashed curves correspond to the cases when it equals to $10\vert\alpha\vert^2$, $\vert\alpha\vert^2$ and $0.1\vert\alpha\vert^2$,
      respectively.
    The 2$\sigma$ experimental bound on $1/\Gamma(p \to \pi^0 e^+)$ obtained at the Super-Kamiokande~\cite{sk}
     is shown by the solid horizontal line, and the 2$\sigma$ sensitivity expected at the Hyper-Kamiokande~\cite{hk} by the dotted horizontal line.
    The 2$\sigma$ range of the latest result of the top quark mass measurement by the ATLAS Collaboration~\cite{topatlas} is shown by the vertical lines,
     with the solid one corresponding to the central value and the dashed ones to the 2$\sigma$ range.
    }
    \label{mtvsdecayrate}
  \end{center}
\end{figure}

We numerically derive the correlation between the proton decay partial width $\Gamma(p \to \pi^0 e^+)$ and the top quark pole mass $m_t^{{\rm pole}}$.
The parameters in the proton decay partial width are set according to Ref.~\cite{nath} as $D+F=1.267$ and $\vert \alpha \vert=\vert \beta \vert=0.009$~GeV$^3$.
The two-loop SM RG equations in Ref.~\cite{2looprge} are used to evaluate $\Lambda_{cr}$, by fixing the Higgs boson mass at $m_h=125.09$~GeV,
 the $W$ boson mass at $M_W=80.384$~GeV and the strong gauge coupling at the $Z$ boson pole at $\alpha_S(M_Z)=0.1184$,
 while variating the top quark pole mass.
The RG running of the Higgs quartic coupling is shown in Figure~\ref{rge}
 for three representative cases with $m_t^{{\rm pole}}=171.44$~GeV, $172.84$~GeV and $174.24$~GeV.
These values are cited from the 2$\sigma$ range of the combined result of the top quark mass measurement by the ATLAS Collaboration~\cite{topatlas}.
Since the coupling constants $h_1,h_2,h_3,h_4,h_5$ are $O(1)$ but unknown, we variate 
 $(\vert \beta h_1 + \beta h_2 + \alpha h_4 \vert^2 + \vert \alpha h_3 + \beta h_5 \vert^2)$ from $10\vert\alpha\vert^2$ to $0.1\vert\alpha\vert^2$.
The result is presented in Figure~\ref{mtvsdecayrate}, where the solid curve corresponds to the case when 
 $\vert \beta h_1 + \beta h_2 + \alpha h_4 \vert^2 + \vert \alpha h_3 + \beta h_5 \vert^2 = \vert\alpha\vert^2$,
 and the lower and upper dashed curves, respectively, correspond to the cases when it equals to $10\vert\alpha\vert^2$ and $0.1\vert\alpha\vert^2$.
Also shown are the current 2$\sigma$ experimental bound on the proton decay partial width obtained at the Super-Kamiokande~\cite{sk},
 $1/\Gamma(p \to \pi^0 e^+) > 1.4\times10^{34}$~yrs, denoted by the solid horizontal line,
 and the 2$\sigma$ sensitivity expected at the Hyper-Kamiokande~\cite{hk} with a 5.6 Megaton$\cdot$year exposure, 
 $1/\Gamma(p \to \pi^0 e^+) > 1.3\times10^{35}$~yrs, denoted by the dotted horizontal line.
As a reference, we display the 2$\sigma$ range of the latest combined result of the top quark mass measurement by the ATLAS Collaboration~\cite{topatlas},
 which has reported $m_t=172.84\pm0.70$~GeV,
 by the vertical lines, with the solid one corresponding to the central value and the dashed ones to the 2$\sigma$ range.
The CMS Collaboration has reported a consistent result~\cite{topcms}.
Note that the ATLAS Collaboration has also conducted the determination of the top quark pole mass
 by employing the differential cross section for the production of a top quark pair + 1-jet and has reported
 $m_t^{{\rm pole}}=173.7-2.1+2.3$~GeV~\cite{pole}, in agreement with the corresponding CMS result~\cite{polecms}.
The figure tells that if future determinations of the top quark pole mass yield a value above $\sim 172.5$~GeV, we have a chance to
 observe $p \to \pi^0 e^+$ events at the Hyper-Kamiokande.

To summarize, we have studied a scenario based on gauge-Higgs unification where the scale at which the Higgs quartic coupling vanishes in the SM corresponds to
 the KK scale of the 5D compactified spacetime.
The KK scale is related with the 5D Planck scale.
Since the 1st generation fermions are mostly localized at an orbifold fixed point,
 quantum gravity can give rise to operators involving four 1st generation fermions suppressed by the square of the 5D Planck scale.
Hence, the 5D Planck scale, or equivalently the KK scale, determines the partial width of the $p \to \pi^0 e^+$ process induced by 5D Planck suppressed operators.
We have thus obtained a correlation between the top quark mass, which controls the RG running of the Higgs quartic coupling, and the proton partial decay width.
The correlation indicates that the future Hyper-Kamiokande experiment may discover the proton decay if the top quark pole mass is larger than about 172.5~GeV.
\\

\section*{Acknowledgement}

This work is partially supported by Scientific Grants by the Ministry of Education, Culture, Sports, Science and Technology of Japan (Nos. 24540272, 26247038, 15H01037, 16H00871, and 16H02189)
 and the United States Department of Energy (DE-SC 0013680).


\begin{thebibliography}{99}
 \bibitem{higgs}
    G.~Aad {\it et al.} [ATLAS and CMS Collaborations],
  %``Combined Measurement of the Higgs Boson Mass in $pp$ Collisions at $\sqrt{s}=7$ and 8 TeV with the ATLAS and CMS Experiments,''
  Phys.\ Rev.\ Lett.\  {\bf 114}, 191803 (2015)
  %doi:10.1103/PhysRevLett.114.191803
  [arXiv:1503.07589 [hep-ex]].
 
 \bibitem{topatlas}
   M.~Aaboud {\it et al.} [ATLAS Collaboration],
  %``Measurement of the top quark mass in the $t\bar{t}\to$ dilepton channel from $\sqrt{s}=8$ TeV ATLAS data,''
  arXiv:1606.02179 [hep-ex].
 
 \bibitem{topcms}
  V.~Khachatryan {\it et al.} [CMS Collaboration],
  %``Measurement of the top quark mass using charged particles in pp collisions at $\sqrt s =$ 8 TeV,''
  Phys.\ Rev.\ D {\bf 93}, no. 9, 092006 (2016)
  %doi:10.1103/PhysRevD.93.092006
  [arXiv:1603.06536 [hep-ex]].
 
 
 \bibitem{lifetime}
    J.~Elias-Miro, J.~R.~Espinosa, G.~F.~Giudice, G.~Isidori, A.~Riotto and A.~Strumia,
  %``Higgs mass implications on the stability of the electroweak vacuum,''
  Phys.\ Lett.\ B {\bf 709}, 222 (2012)
  %doi:10.1016/j.physletb.2012.02.013
  [arXiv:1112.3022 [hep-ph]].
 
 
 \bibitem{ghu}
   N.~S.~Manton,
  %``A New Six-Dimensional Approach To The Weinberg-Salam Model,''
  Nucl.\ Phys.\ B {\bf 158}, 141 (1979);
  D.~B.~Fairlie,
  %``Higgs' Fields And The Determination Of The Weinberg Angle,''
  Phys.\ Lett.\ B {\bf 82}, 97 (1979); 
  %``Two Consistent Calculations Of The Weinberg Angle,''
  J.\ Phys.\ G {\bf 5}, L55 (1979);
  Y.~Hosotani,
  %``Dynamical Mass Generation By Compact Extra Dimensions,''
  Phys.\ Lett.\ B {\bf 126}, 309 (1983), 
  %``Dynamical Gauge Symmetry Breaking As The Casimir Effect,''
  Phys.\ Lett.\ B {\bf 129}, 193 (1983), 
  %``DYNAMICS OF NONINTEGRABLE PHASES AND GAUGE SYMMETRY BREAKING,''
   Annals Phys.\  {\bf 190}, 233 (1989).
 


 \bibitem{eft}
   N.~Haba, S.~Matsumoto, N.~Okada and T.~Yamashita,
  %``Effective theoretical approach of Gauge-Higgs unification model and its phenomenological applications,''
  JHEP {\bf 0602}, 073 (2006)
  %doi:10.1088/1126-6708/2006/02/073
  [hep-ph/0511046].
  
  
 \bibitem{scrucca}
    C.~A.~Scrucca, M.~Serone, L.~Silvestrini and A.~Wulzer,
  %``Gauge Higgs unification in orbifold models,''
  JHEP {\bf 0402}, 049 (2004)
  %doi:10.1088/1126-6708/2004/02/049
  [hep-th/0312267].
  
  
  \bibitem{csaki}
   G.~Cacciapaglia, C.~Csaki and S.~C.~Park,
  %``Fully radiative electroweak symmetry breaking,''
  JHEP {\bf 0603}, 099 (2006)
  %doi:10.1088/1126-6708/2006/03/099
  [hep-ph/0510366].
  
   
 \bibitem{nath}
   P.~Nath and P.~Fileviez Perez,
  %``Proton stability in grand unified theories, in strings and in branes,''
  Phys.\ Rept.\  {\bf 441}, 191 (2007)
  %doi:10.1016/j.physrep.2007.02.010
  [hep-ph/0601023].
  
 
 \bibitem{2looprge}
   D.~Buttazzo, G.~Degrassi, P.~P.~Giardino, G.~F.~Giudice, F.~Sala, A.~Salvio and A.~Strumia,
  %``Investigating the near-criticality of the Higgs boson,''
  JHEP {\bf 1312}, 089 (2013)
  %doi:10.1007/JHEP12(2013)089
  [arXiv:1307.3536 [hep-ph]].

 
 \bibitem{pole}
     G.~Aad {\it et al.} [ATLAS Collaboration],
  %``Determination of the top-quark pole mass using $ t\overline{t} $ + 1-jet events collected with the ATLAS experiment in 7 TeV pp collisions,''
  JHEP {\bf 1510}, 121 (2015)
  %doi:10.1007/JHEP10(2015)121
  [arXiv:1507.01769 [hep-ex]].
 
 
 \bibitem{sk}
  M.~Miura [Super-Kamiokande Collaboration],
  %``Search for Nucleon Decay in Super-Kamiokande,''
  Nucl.\ Part.\ Phys.\ Proc.\  {\bf 273-275}, 516 (2016).
  
 \bibitem{hk}
    K.~Abe {\it et al.},
  %``Letter of Intent: The Hyper-Kamiokande Experiment --- Detector Design and Physics Potential ---,''
  arXiv:1109.3262 [hep-ex].
 
 \bibitem{polecms}
  S.~Chatrchyan {\it et al.} [CMS Collaboration],
  %``Determination of the top-quark pole mass and strong coupling constant from the t t-bar production cross section in pp collisions at $\sqrt{s}$ = 7 TeV,''
  Phys.\ Lett.\ B {\bf 728}, 496 (2014)
  Erratum: [Phys.\ Lett.\ B {\bf 738}, 526 (2014)]
  %doi:10.1016/j.physletb.2014.08.040, 10.1016/j.physletb.2013.12.009
  [arXiv:1307.1907 [hep-ex]].
  
\end{thebibliography}
\end{document}